  \documentstyle[prd, preprint,eqsecnum,aps]{revtex}  
  \begin{document}
  \draft
   \preprint{\vbox{
   \hbox{UMTH-00-06}
   \hbox{hep-th/0003132}
  }}

  \title{On the Relationship of Gravitational Constants in KK Reduction\\ }

  \author {J. X. Lu}
  \address{Randall Physics Laboratory
  University of Michigan, Ann Arbor, MI 48109-1120\\
   E-mails: jxlu@umich.edu}

  \maketitle
  
  \begin{abstract}
  In this short note, we try to clarify a seemly trivial but often
  confusing question in relating a higher-dimensional 
  {\it physical} gravitational constant to its lower-dimensional correspondence 
  in Kaluza-Klein reduction. In particular, we re-derive 
  the low-energy M-theory gravitational constant in
  terms of type IIA string coupling $g_s$ and constant $\alpha'$
   through the metric relation between the two theories.   
  \end{abstract}
  \newpage
  
The proper determination of eleven dimensional M-theory gravitational 
constant (therefore, the eleven-dimensional Planck constant), 
in terms of type IIA string coupling $g_s$ and constant $\alpha'$, 
is important, for example, for the BFSS matrix proposal of 
M-theory\cite{banfss}. It is also important for whether brane modes can
possibly decouple from bulk gravity modes \cite{ahagmoo,aliosj,lur} 
in the so-called decoupling limit. Given the string constant $\alpha'$ (
therefore the units in type IIA string theory) and the relationship
between 11-D M-theory and type IIA string theory, the 11-D
M-theory {\it physical } gravitational constant as well as the units for
M-theory are also given. We therefore expect a precise expression for
the M-theory gravitational constant in terms of type IIA string coupling
$g_s$ and constant $\alpha'$. However, there exist no unique answers in the 
literature for this constant. We try to clarify, in this short note,        
possible confusion about the derivations of this constant.
 
 Let us begin with a general discussion in relating a higher-dimensional
physical gravitational constant to its lower-dimensional correspondence 
in dimension redection. We start with the following gravity action in 
dimension D 
\begin{equation}
I_D = \frac{1}{2 \bar{\kappa}^2_D} \int d^D x \sqrt{- \det \hat G} \left[{\hat R} +
\cdots\right],
\label{eq:hda}
\end{equation}   
where ${\hat G}_{MN}$ is the metric, $\hat R$ is the Ricci scalar, 
the constant ${\bar \kappa}^2_D$ is usually called gravitational constant (the
Newton constant $8\pi G_N^D \equiv  \kappa^2_D$)\footnote{The $\kappa$
without a bar  corresponds to the physical gravitional constant while
the one with a bar is not necesarily a physical one, see the
explanation given in the text.},  and
$\cdots$ in the above action represents other possible
fields\footnote{For the purpose of this paper, we  need to consider
only the first term in the action.}. 

Now we wish to compactify the above action to dimension d ($< D$). For our
purpose, we need to consider only the massless graviton whose effective
action is
\begin{equation}
I_d = \frac{1}{2 {\bar \kappa}_d^2} \int d^d x \sqrt{- \det g}\left[R +
\cdots\right].
\label{eq:lda}
\end{equation}

In obtaining the above action from Eq.\ (\ref{eq:hda}), we made the
split for the higher-dimensional coordinate 
$x^M = (x^\mu, y^i)$ with $M = 0, 1 \cdots, D - 1; 
\mu = 0, 1, \cdots d - 1$ and $i = 1, \cdots D - d$. Here $x^\mu$ are
the coordinates of the lower-dimensional spacetime and $y^i$ are the
compactified coordinates. We therefore have the same units for both the 
D-dimensional theory and the compactified d-dimensional theory. 
The massless sector of the lower-dimensional
theory can be obtained by assuming the higher-dimensional fields to be
independent of $y^i$. We then simply integrate out the $y^i$ from the
action \ (\ref{eq:hda}). By comparing the resulted action with 
Eq.\ (\ref{eq:lda}), we have the relation for the gravitational
constants and the compactified volume $V_{D - d}$ as\footnote{I will
make this precise when we discuss on how to  reduce the low energy
M-theory to the  low energy type IIA  string theory.}
\begin{equation}
\kappa^2_D = \kappa_d^2\, V_{D - d}.
\label{eq:gcr}
\end{equation} 
We often say that the above equation, relating the higher-dimensional
physical gravitational constant to its lower-dimensional correspondence
through the physical volume measured with respect to the
lower-dimensional metric, is independent of the actual metric relation
between the higher-dimensional and the lower-dimensional theories. This 
is true indeed. 

However, what is often ignored in practice is the
implicit assumption used in deriving Eq.\ (\ref{eq:gcr}) that we choose 
the asymptotic metric (i.e., the underlying vacuum) for the
higher-dimensional theory to be same as that for the compactified
lower-dimensional theory\footnote{Usually we choose the
asymptotic metric for the lower-dimensional theory to be Minkowskian,
i.e., flat Minkowski metric $\eta_{\mu\nu} = (-, +, \cdots, +)$, see the
definition given in \cite{polone}. This is the metric used in defining
the {\it physical} gravitational constant. This is also the metric used in
perturbative string theory in defining the string tension $T_f = 1/(2\pi
\alpha')$ or the string constant $\alpha'$.}. Only in this
case, we can take the constant 
${\bar \kappa}^2$ in front of the respective action as the physical
gravitational constant\footnote{In this case, for example, the metric 
$g_{\mu\nu} = \eta_{\mu\nu} +\kappa h_{\mu\nu}$ for small graviton fluctuation
around flat Minkowski spacetime and the Einstein-Hilbert action reduces
to the cannonical form $\int (\partial h)^2 + \kappa (\partial h)^2 h$.}. In
general, however, the asymptotic metrics for both the 
higher-dimensional theory and the compactified lower-dimensional theory
are not necessarily the same because the scalars due to the compactification
 develop VEV \footnote{Of course, one can always define both 
of the higher-dimensional metric and the lower-dimensional one to be
same asymptotically by absorbing the possible constant factor due to the
VEV of scalars into the ${\bar \kappa}^2$ in front of the action. For
the higher-dimensional theory, the compactified coordinates $y^i$ have
to be rescaled properly with respect to the lower-dimensional metric,
see the example given later in relating M-theory to type IIA string. Then
the resulting $\kappa^2$ is the physical one. This is what Maldacena did
in \cite{mal} for  obtaining masses properly for BPS
states in string theory using U-duality.}. If this happens, we cannot take both 
${\bar \kappa}_D^2$ and ${\bar \kappa}_d^2$ in front of the respective 
action in the above as physical\footnote{Polchinski in \cite{poltwo}
chose both the units and metric for M-theory the same as those for
string theory. By definition, his $\kappa^2_{11}$ and the compactified 
radius are both physical, not the usual ${\bar \kappa}^2_{11}$ and the coordiante
radius $r$.}. The ignorance of this fact is often the 
source of confusion in the literature. For example, the low-energy M-theory
physical gravitational constant has been given correctly  
in \cite{banfss,polone} as $2 \kappa^2_{11} = (2 \pi)^8 g_s^3
\alpha'^{9/2}$ in terms of type IIA string coupling $g_s$ and constant 
$\alpha'$. But this constant has also been given incorrectly 
in the literature precisely because ${\bar \kappa}$ is mistaken as
$\kappa$.
    
	In the remainder of this note, I will focus, as an example, on  
the reduction of 11-dimensional low-energy M-theory on a circle $S^1$ to
give the low energy theory of type IIA string. The 11-D M-theory metric 
is related to the type IIA string metric as
\begin{equation}
d s^2_{11} = e^{- 2 \phi /3} d s^2_{10} + e^{4 \phi /3} (d x^{11})^2,
\label{eq:mr}
\end{equation}
where $\phi$ is the dilaton in type IIA string theory, 11-th coordinate $x^{11}$
has a period $2 \pi r$ with the coordinate radius $r$. In the above we
have dropped the KK
vector field $A_{\mu}$. For our purpose, $A_\mu$ is
irrelevant. We take type IIA string metric 
$d s^2_{10}$ as asympotically Minkowski since that is where we quantize
the perturbative type IIA string. That is the metric used in defining
the fundamental string tension $T_f = 1/(2\pi \alpha')$. So are the tensions for
D-branes and NSNS branes. With Eq.\ (\ref{eq:mr}) and taking $D
= 11$ in Eq.\ (\ref{eq:hda}), we have the low energy action of type IIA
string as
\begin{equation}
I_{10} = \frac{2\pi r}{\bar{\kappa}_{11}^2} \int d^{10} x \sqrt{-\det g}
e^{- 2\phi} \left[R + \cdots \right],
\label{eq:10a}
\end{equation}
where $g_{\mu\nu}$ is the string metric. By definition, we have the
following relation 
\begin{equation}
{\bar\kappa}_{11}^2 = 2\pi r\, e^{- 2 \phi_0}\, \kappa_{10}^2,
\label{eq:nr}
\end{equation}
where $\kappa_{10}^2 \equiv {\bar\kappa}_{10}^2\, e^{2 \phi_0}$ is the
physical gravitational constant in D = 10. $\phi_0$ is the VEV of
dilaton or the asymptotic value of the dilaton and is related to the
string coupling as $g_s = e^{\phi_0}$. 

	As we stress above that Eq.\ (\ref{eq:gcr}) holds true always. In 
the present context, it is 
\begin{equation}
\kappa_{11}^2 = 2\pi\rho \,\kappa_{10}^2,
\label{eq:pnr}
\end{equation}
with $\rho$ the physical radius. 
Let me explain why $2 \kappa^2_{11} = (2 \pi)^8 g_s^3
\alpha'^{9/2}$  given in \cite{banfss,polone} must be correct. 
As I mentioned above, 
Eq.\ (\ref{eq:pnr}) should hold always true. The physical gravitational
constant $2\kappa_{10}^2 = (2 \pi)^7 g_s^2 \alpha'^4 $ was given  
in \cite{dea}. As we now know that the strong coupling of type IIA
string is just M-theory compactified on a big circle. In order for this
to be true, one needs to identify the spectrum of D0 branes 
with that of  momentum (Kaluza-Klein) states. 
This implies that the physical radius of the
circle measured in string metric is given as the inverse of mass of a
single D0 brane, i.e., $\rho = g_s \alpha'^{1/2}$. 
Then we have the 11-D physical 
gravitational constant from Eq.\ (\ref{eq:pnr}) as given above.

From Eqs.\ (\ref{eq:nr}) and (\ref{eq:pnr}), we have 
\begin{equation}
\frac{{\bar\kappa}_{11}^2} {r\, e^{- 2\phi_0}} = \frac{\kappa_{11}^2} {\rho}.
\label{eq:rr}
\end{equation}
We intend to determine the relation
between $\kappa_{11}$ and ${\bar \kappa}_{11}$ and the relation between
$r$ and $\rho$ unambiguously. To my knowledge, no explicit derivations
for these two relations have been given in the literature. We dare to
present them here.

	For our purpose, we need to consider only the asymptotic
metric relation in Eq.\ (\ref{eq:mr}), i.e., 
\begin{eqnarray}
(d s_0)^2_{11}  &=& e^{- 2 \phi_0 /3} (d s_0)^2_{10}  + e^{4
\phi_0 / 3} (d x^{11})^2,\nonumber\\
&=& e^{- 2\phi_0 / 3} \left[ (d s_0)^2_{10}  + (d {\tilde x}^{11})^2\right],
\label{eq:asymr}
\end{eqnarray}
where the asymptotic string metric $(d s_0)^2_{10} $ is actually 
Minkowskian and the rescaled 11-th coordinate ${\tilde x}^{11} =
e^{\phi_0} x^{11}$ with its radius $\tilde r = e^{\phi_0} r$. The first
line in the above equation indicates clearly that the 11-D metric cannot
be asymptotically Minkowskian if we insist $d s^2_{10}$ be
so\footnote{We can no longer rescale $(d s_0)^2_{10}$ since that is the
metric used in defining the string constant $\alpha'$.}. The second
line says that the 11-D metric can be made asymptotically Minkowskian
 up to a constant scaling factor $e^{- 2 \phi_0 /3}$ if we rescale $x^{11}$ to $
{\tilde x}^{11}$ as given above. In other words, the scaled radius
$\tilde r$ is measured with respect to the string metric. Because the
string constant $\alpha'$ is defined with respect to the string metric,
the second line in Eq.\ (\ref{eq:asymr}) should be used in the following 
equation.  

	By definition, from the above asymptotic metric relation, we
 have
\begin{equation}
\kappa_{11}^2 = \frac{{\bar \kappa}_{11}^2} { 
\sqrt{-\det \hat G_0} {\hat G_0}^{- 1}} = e^{3 \phi_0}\, {\bar
\kappa}_{11}^2,
\label{eq:pugcr}
\end{equation}
where ${\hat G}_0$ denotes the 11-D asymptotic metric given in the
second line of Eq.\ (\ref{eq:asymr}). Using the above and 
Eq.\ (\ref{eq:rr}), we  derive $r = \alpha'^{1/2}$. Then $\tilde r =
e^{\phi_0}\, \alpha'^{1/2} = \rho$ is the physical radius measured in the
string metric. 

	Let us provide an independent check of Eq.\ (\ref{eq:pugcr}).
For simplicity, we consider the reduction of a scalar field $\Phi (x^M)$ 
from 11-D to 10-D on a circle $S^1$. The usual KK reduction says
\begin{equation}
\Phi (x^\mu, x^{11}) = \sum_{n = -\infty}^\infty \Phi_n (x^\mu) e^{i n
x^{11} / r},
\label{eq:skkr}
\end{equation}
or 
\begin{equation}
\Phi (x^\mu, {\tilde x}^{11}) = \sum_{n = -\infty}^\infty \Phi_n (x^\mu) 
e^{i n {\tilde x}^{11} / {\tilde r}},
\label{eq:skk}
\end{equation}
where $x^{11}$ or ${\tilde x}^{11}$ is
the compactified coordinate defined earlier. The 11-D wave equation 
$\bigtriangledown_M \bigtriangledown^M \Phi = 0$ in the asymptotic
region (or around the Minkowski vacuum) becomes
\begin{equation}
\eta^{\mu\nu} \partial_\mu \partial_\nu \Phi_n (x^\mu) = 
\frac{n^2}{r^2 e^{2\phi_0}} \Phi_n (x^\mu),
\label{eq:sspr}
\end{equation}
where we have used the first line in Eq.\ (\ref{eq:asymr}), or we have
\begin{equation}
\eta^{\mu\nu} \partial_\mu \partial_\nu \Phi_n (x^\mu) = 
\frac{n^2}{{\tilde r}^2} \Phi_n (x^\mu),
\label{eq:ssp}
\end{equation} 
where we have used the second line in Eq.\ (\ref{eq:asymr}). The mass
spectrum with respect to the 10-D Minkowski vacuum in string frame
can be obtained from either of the above equations as
\begin{equation}
M_n^2 = - p_\mu p^\mu = \frac{n^2}{r^2 e^{2\phi_0}} = \frac{n^2}{{\tilde r}^2}.
\label{eq:msp}    
\end{equation}
From the above, we should identify $\tilde r = r e^{\phi_0}$ as the 
physical radius with respect to the string metric. The mass $M_n$ should be
identified with that of  n D0 brane for the reason mentioned earlier.
We therefore have $\tilde r = g_s
\alpha'^{1/2}$. So we have $r = \alpha'^{1/2}$. Using Eq.\ (\ref{eq:rr})
and $\rho = \tilde r$, we obtain also Eq.\ (\ref{eq:pugcr}). Our
discussion gives $2 {\bar \kappa}^2_{11} = (2 \pi)^8 \alpha'^{9/2}$.  

	In summary, we explain how to obtain the physical gravitational
constant for the original higher dimensional theory if we know the
physical gravitational constant in the compactified lower-dimensional theory.
In particular, we derive the relation 
$\kappa_{11}^2 = g_s^3 {\bar \kappa}_{11}^2$ and determine the radius $r
= \alpha'^{1/2}$ (or $\tilde r = g_s \alpha'^{1/2}$)
unambiguously.

 \acknowledgments 
The author would like to thank Dan Chung for discussion and
to acknowledge the support of U. S. Department of Energy.

  \end{document}